\documentclass{article}

\usepackage{PRIMEarxiv}

\usepackage[utf8]{inputenc} % allow utf-8 input
\usepackage[T1]{fontenc}    % use 8-bit T1 fonts
\usepackage{url}            % simple URL typesetting
\usepackage{booktabs}       % professional-quality tables
\usepackage{amsfonts}       % blackboard math symbols
\usepackage{amssymb}        % symbols like \bigstar (loaded once only)
\usepackage{nicefrac}       % compact symbols for 1/2, etc.
\usepackage{microtype}      % microtypography
\usepackage{lipsum}
\usepackage{graphicx}       % graphics
\usepackage{float}
\usepackage{pifont}         % for \ding{}
\usepackage[font=it, labelfont=bf, justification=centering]{caption}
\usepackage{subcaption}
\usepackage[version=4]{mhchem}  % for chemical equations
\graphicspath{{media/}}     % organize images under media/

\usepackage{hyperref}       % hyperlinks
\usepackage[table,dvipsnames]{xcolor}
\hypersetup{
    colorlinks=true,
    linkcolor=blue!60!black,  % dark blue for internal links (e.g., figures, sections)
    citecolor=blue!50!black,  % dark blue for citations
    urlcolor=blue!30!black    % slightly lighter for URLs
}

\usepackage{fancyhdr}
\title{Bubble-Burst Synthesis of Ammonia, Amino Acids, and Urea Under Ambient, Catalyst-Free Conditions
}

\author{
  \href{https://mse.ncsu.edu/people/cuomo/}{Jerome J. Cuomo}\quad
  \href{https://www.linkedin.com/in/ian-goodall-aaaa17189/}{Ian Goodall}\quad
  \href{https://mse.ncsu.edu/people/rguarni/}{C. Richard Guarnieri}\quad
  \href{https://www.linkedin.com/in/stephenhudak/}{Stephen Hudak} \\[1ex]
  \textit{Department of Materials Science and Engineering} \\
  \textit{North Carolina State University} \\
  \textit{Raleigh, NC, USA} \\[2ex]
  \href{https://en.wikipedia.org/wiki/Jerry_Cuomo}{Gennaro (Jerry) Cuomo} \\[0.5ex]
  \textit{Wild Ducks LLC} \\
  \textit{\href{https://wildducks.us}{wildducks.us}}
}

\begin{document}
\maketitle

\begin{abstract}
This study introduces a catalyst-free, ambient-temperature method for synthesizing nitrogen-based compounds critical to fertilizer production—namely ammonia, urea, ammonium salts, and amino acids. The process relies on bubble-burst-induced microenvironments, where gas bubbles undergo rapid growth and collapse, releasing intense localized energy sufficient to dissociate nitrogen and water molecules. These high-energy zones produce reactive species, including atomic hydrogen (H\textsuperscript{*}) that facilitate nitrogen fixation and drive subsequent chemical transformations without needing catalysts or elevated conditions. Experimental validation using colorimetric assays, Raman spectroscopy, and microscopy confirms the \textit{in situ} formation of ammonia and its downstream conversion into structurally relevant compounds, including peptide-like assemblies. The system supports reaction tuning through dissolved carbon dioxide (CO\textsubscript{2}) or organic acids and can be enhanced by low-energy inputs such as UV or ultrasound. Its simplicity, modularity, and ability to operate without external infrastructure offer a practical and scalable platform for decentralized fertilizer generation and sustainable biochemical production.
\end{abstract}


\section{Introduction}

Nitrogen-based compounds derived from or built upon ammonia are fundamental to fertilizer production, pharmaceuticals, and industrial manufacturing. They also support global food production, medical advancements, and chemical synthesis. Their large-scale production has long relied on the \href{https://www.britannica.com/technology/Haber-Bosch-process}{Haber--Bosch process}, which converts nitrogen (N\textsubscript{2}) and hydrogen (H\textsubscript{2}) into ammonia (NH\textsubscript{3}) under extreme temperatures and pressures~\cite{ref01_hager2008,ref02_smil2001}. While revolutionary, this method remains one of the most energy-intensive industrial processes, consuming vast fossil fuel resources and accounting for over 1.5\% of global carbon dioxide emissions. As demand rises, the need for sustainable, low-energy alternatives has never been more pressing~\cite{ref03_erisman2008}.

Alternative methods such as microbial fermentation and chemical catalysis offer promising pathways for nitrogen-based compound synthesis, but face significant limitations in scalability, cost, and flexibility. Fermentative processes, though biologically efficient, suffer from slow reaction rates and strict substrate requirements~\cite{ref02_smil2001}, while catalytic systems depend on expensive or scarce metals like ruthenium and iridium~\cite{ref05_perez2020}. Other emerging techniques explore self-sustaining reaction conditions, contributing valuable theoretical insights, though practical implementation remains in its early stages. Many of these methods also lack the adaptability required for decentralized, small-scale production, further constraining their accessibility and long-term sustainability.

This study aims to address these issues by exploring a catalyst-free method for synthesizing nitrogen-based compounds—including urea, amino acids, and ammonium salts—under ambient conditions. These conditions refer to standard atmospheric pressure and near-room temperatures. In this process, gas is introduced into an aqueous medium through a bubble-diffusing element, starting with water and a nitrogen-containing gas, such as air or molecular nitrogen (\ce{N2}). Optional additions to the solution, including dissolved CO\textsubscript{2}, simple organic acids, or inorganic anions, enable specific product pathways. These may be further enhanced by low-intensity inputs such as ultraviolet light, ultrasonic agitation, or shear mixing, improving yield and selectivity.

The observation of urea, amino acids, and ammonium salts confirms that ammonia must be formed by the bubble-bursting process, acting as a reactive intermediate. This has been validated through colorimetric tests (e.g., \href{https://www.sciencedirect.com/topics/chemistry/ninhydrin}{Ruhemann’s purple} for amino acids), \href{https://www.edinst.com/resource/what-is-raman-spectroscopy/}{Raman spectroscopy} to confirm urea synthesis, and microscopy to reveal peptide-like assemblies.

Further analysis suggests that the collapse and rupture of microbubbles at the gas–liquid interface generate transient, high-energy microenvironments capable of initiating molecular dissociation. The ambient nature of the process points toward an underlying driving force enabling such transformations. The formation of reactive species, most notably atomic hydrogen, is believed to play a central role, providing the necessary energy for nitrogen activation and enabling the direct, in-medium synthesis of ammonia and its downstream compounds.

Bubble-burst synthesis offers a practical and scalable approach to synthesizing nitrogen-based compounds in aqueous environments. Operating under ambient conditions without specialized infrastructure enables on-site production with minimal energy use and environmental impact. The simplicity and adaptability of this process make it ideal for large-scale and small-scale deployment, supporting applications such as localized fertilizer generation, decentralized biochemical synthesis, and point-of-use production in resource-limited settings.

\section{Methods}
\subsection{System Components and Configuration}

\begin{figure}[H]
  \centering
  \includegraphics[width=0.40\linewidth]{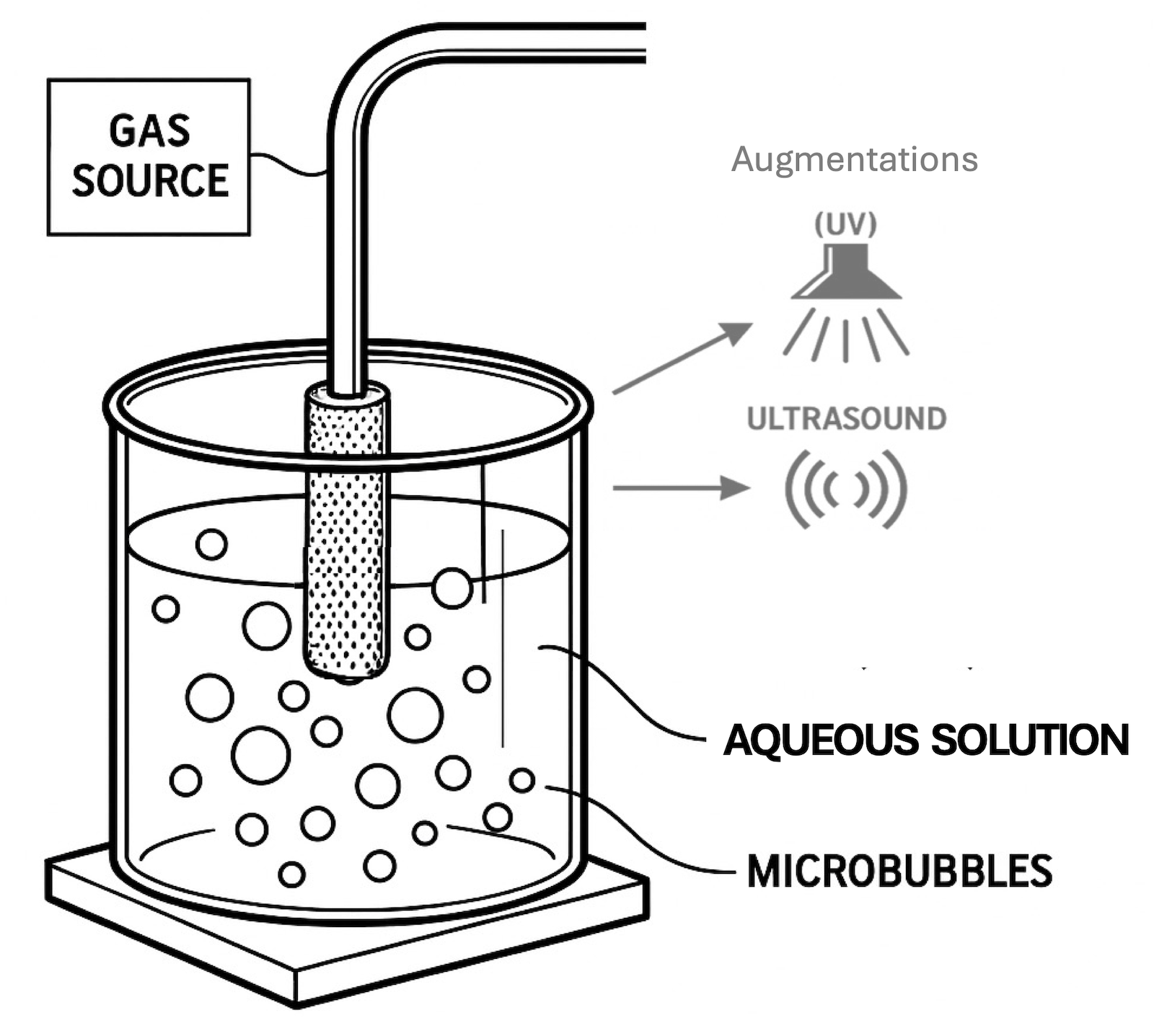}
  \caption{Experimental Setup of the Bubble-Bursting Device.}
  \label{fig:device_setup}
\end{figure}

As illustrated in Figure~\ref{fig:device_setup}, the process begins by introducing a gas feed into a liquid medium reaction system to support continuous gas–liquid interactions. The system includes a gas supply connected to a submerged, porous bubble-diffusing component, typically ceramic or stainless-steel, configured to generate microbubbles within an aqueous medium. These microbubbles propagate throughout the liquid, establishing the reactive interface required for subsequent molecular transformations. 

Our system can be configured for the desired application, such as generating fine, uniformly distributed bubbles for predictable energy release, employing needle-based gas injection for targeted control over bubble frequency and size, or utilizing alternative gas-diffusion systems to adapt to specific reaction conditions. By fine-tuning gas flow parameters, the system maintains consistent cavitation effects, ensuring efficient molecular dissociation and enhancing reaction selectivity.

Beyond standard operation, the system can be enhanced with simple yet effective energy-augmentation components that increase reactivity without the complexity of high-energy infrastructure. For instance, ultrasonic transducers drive violent bubble collapse, focusing energy into micro-zones that enhance radical generation and chemical reactivity~\cite{ref04_lauterborn2010}. UV light sources contribute additional photon energy, promoting bond cleavage and the generation of transient reactive species. High-shear mixing devices further improve gas–liquid interactions by increasing cavitation frequency and ensuring uniform dispersion of reactants, which in turn optimizes energy transfer throughout the reaction medium.

\section{Results}

Experimental results support the \textit{in situ} generation of ammonia and its conversion into various nitrogen-based compounds. Multiple analytical techniques, including colorimetric testing, spectroscopy, and microscopy, confirm these findings.

\subsection{Colorimetric Testing for Amino Acid Presence}

The potential of the bubble-bursting process for amino acid synthesis was evaluated using an aqueous solution of acetic acid and water. Air—composed of approximately 78\% nitrogen, 21\% oxygen, and trace amounts of carbon dioxide (0.04\%)—was bubbled through the solution for several hours using a porous bubbler, creating dynamic gas–liquid interactions. At its core, this experiment tested a fundamental question: Could a process that uses only air and water generate amino acid signatures?

\begin{figure}[H]
  \centering
  \includegraphics[width=0.45\linewidth]{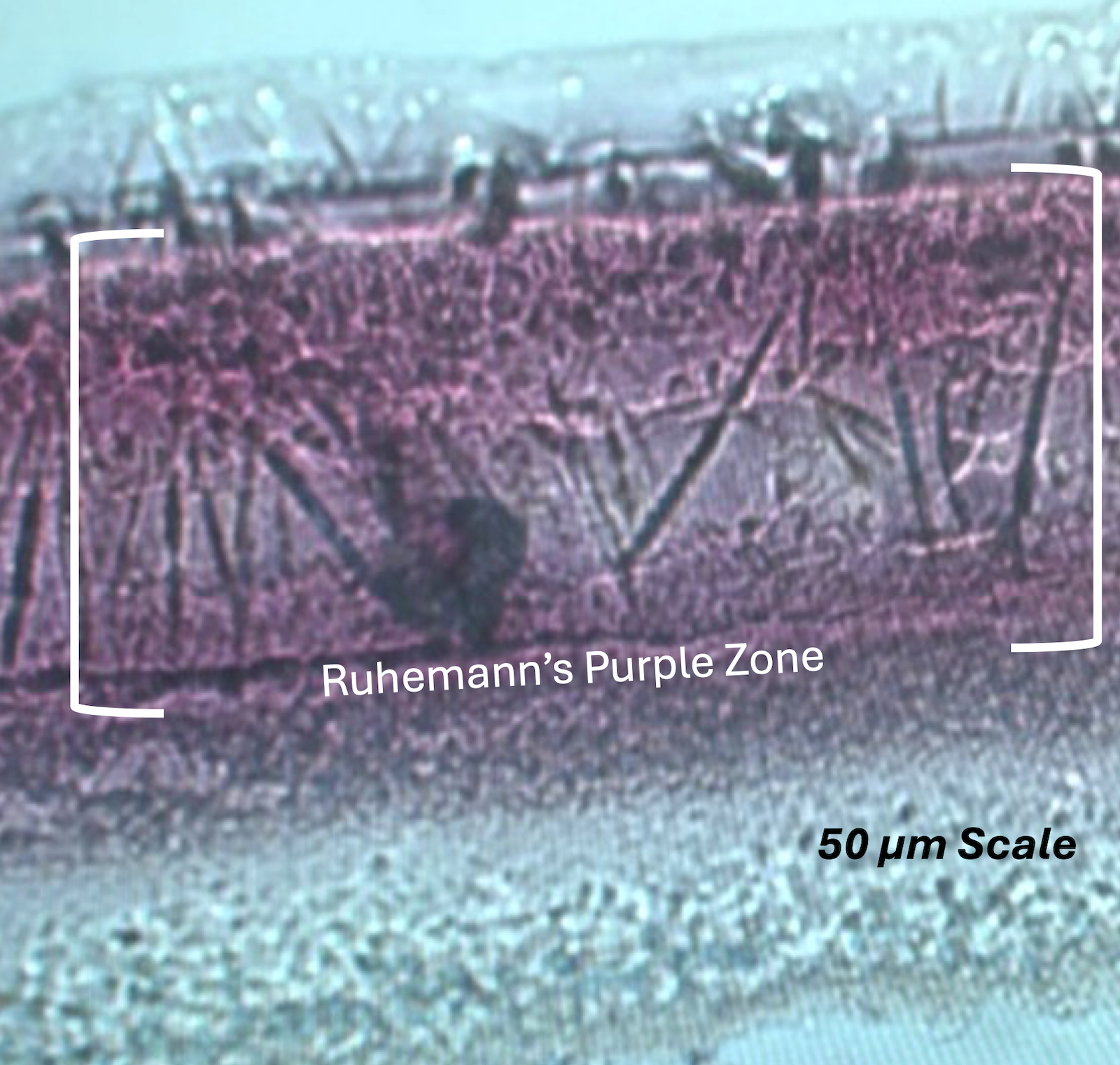}
  \caption{Microscopic Image Showing Ruhemann's Purple\\Indication of Amino Acid Presence.}
  \label{fig:ruhemanns_purple}
\end{figure}

Following the reaction, the solution was subjected to \href{https://www.sciencedirect.com/topics/chemistry/ninhydrin}{ninhydrin testing}, a widely used method for detecting free amino groups. As seen in Figure~\ref{fig:ruhemanns_purple}, the test produced the unmistakable deep purple hue of \href{https://www.researchgate.net/publication/331379104_Ninhydrin_Based_Visible_Spectrophotometric_Determination_of_Gemigliptin}{Ruhemann’s purple}, a well-established indicator of the presence of amino acids. This reaction occurs when ninhydrin interacts with free amino groups, forming a chromophore that serves as an immediate visual confirmation of amino acid formation~\cite{ref06_friedman2004}.

Given the simplicity of the inputs—predominantly nitrogen from air and hydrogen from water—the formation of amino acids strongly suggests ammonia as an intermediate. This finding indicates that the bubble-bursting process not only initiates nitrogen fixation but also generates ammonia as a key intermediate, enabling downstream synthesis of more complex nitrogen-based compounds.

\subsection{Spectroscopy – Urea Evidence}

\begin{figure}[H]
  \centering
  \includegraphics[width=0.6\linewidth]{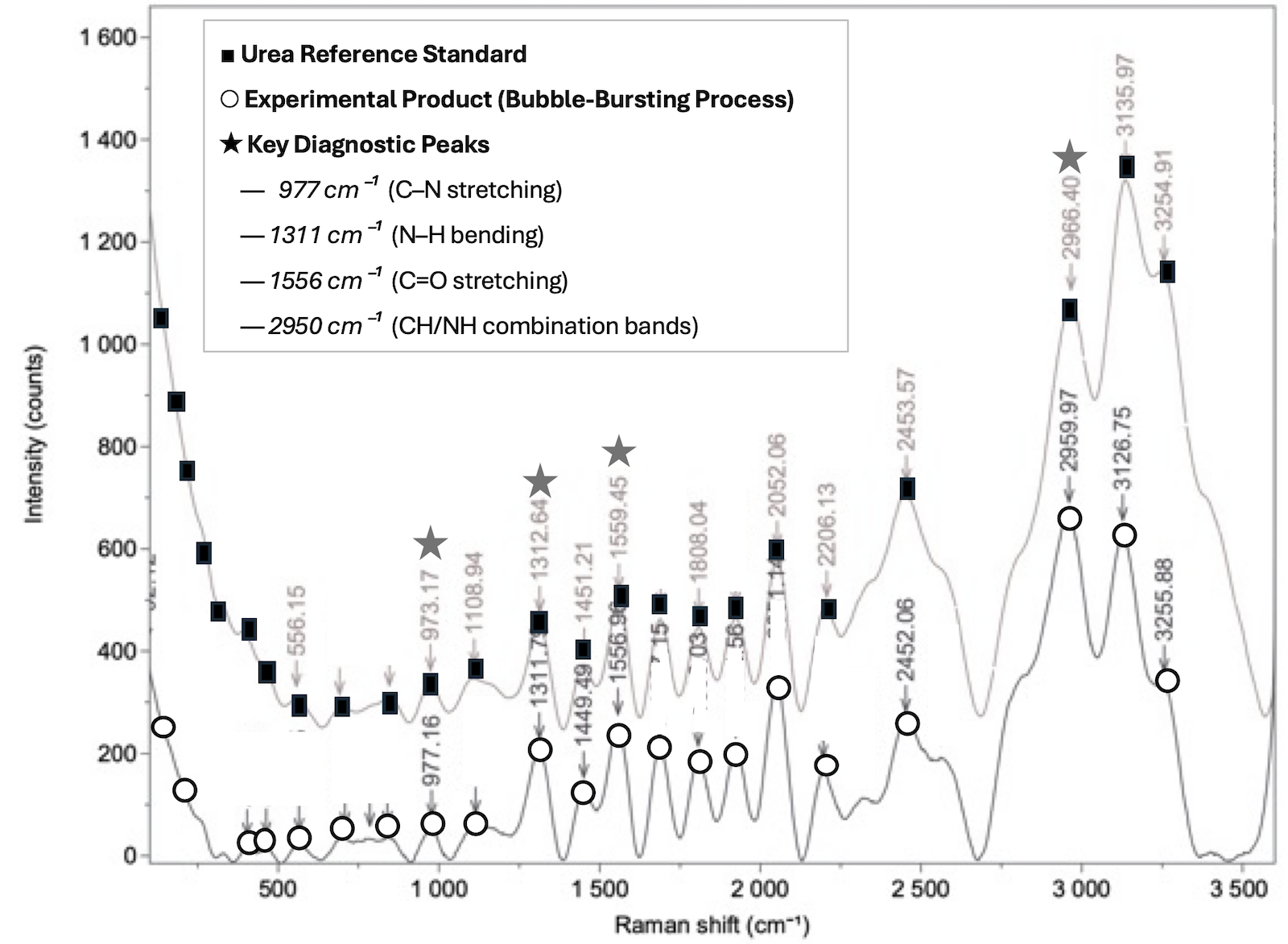}
  \caption{Raman spectra showing urea matches between\\the reference standard and bubble-bursting experiment samples.}
  \label{fig:raman_urea}
\end{figure}

Building on the ninhydrin test results, Raman spectroscopy provided additional validation, confirming the presence of urea, as shown in Figure~\ref{fig:raman_urea}. The top spectrum (marked by \ding{110}, solid square) corresponds to a reference urea standard (99\% purity, \href{https://www.sigmaaldrich.com/US/en/product/sial/u5128}{Sigma-Aldrich}) treated with ninhydrin, while the bottom spectrum (marked by \textopenbullet, hollow circle) represents the experimental product obtained from the bubble-bursting process. Peaks marked by \ding{72} (star symbol) denote the key diagnostic features characteristic of urea.

The spectrum of the experimental residue showed strong correlation with a commercial urea standard, with coincident peaks at approximately 977~cm\textsuperscript{--1} (C–N stretch), 1311~cm\textsuperscript{--1} (N–H bending), and 1556~cm\textsuperscript{--1} (C=O stretch), consistent with the expected vibrational modes of urea. This strong spectral correlation confirms the presence of urea or closely related derivatives in the experimental sample, indicating that the bubble energy can induce urea synthesis under ambient conditions.

\subsection{Microscopic Analysis}

\begin{figure}[htbp]
  \centering

  \begin{subfigure}[t]{0.43\linewidth}
    \centering
    \includegraphics[height=2.5in, keepaspectratio]{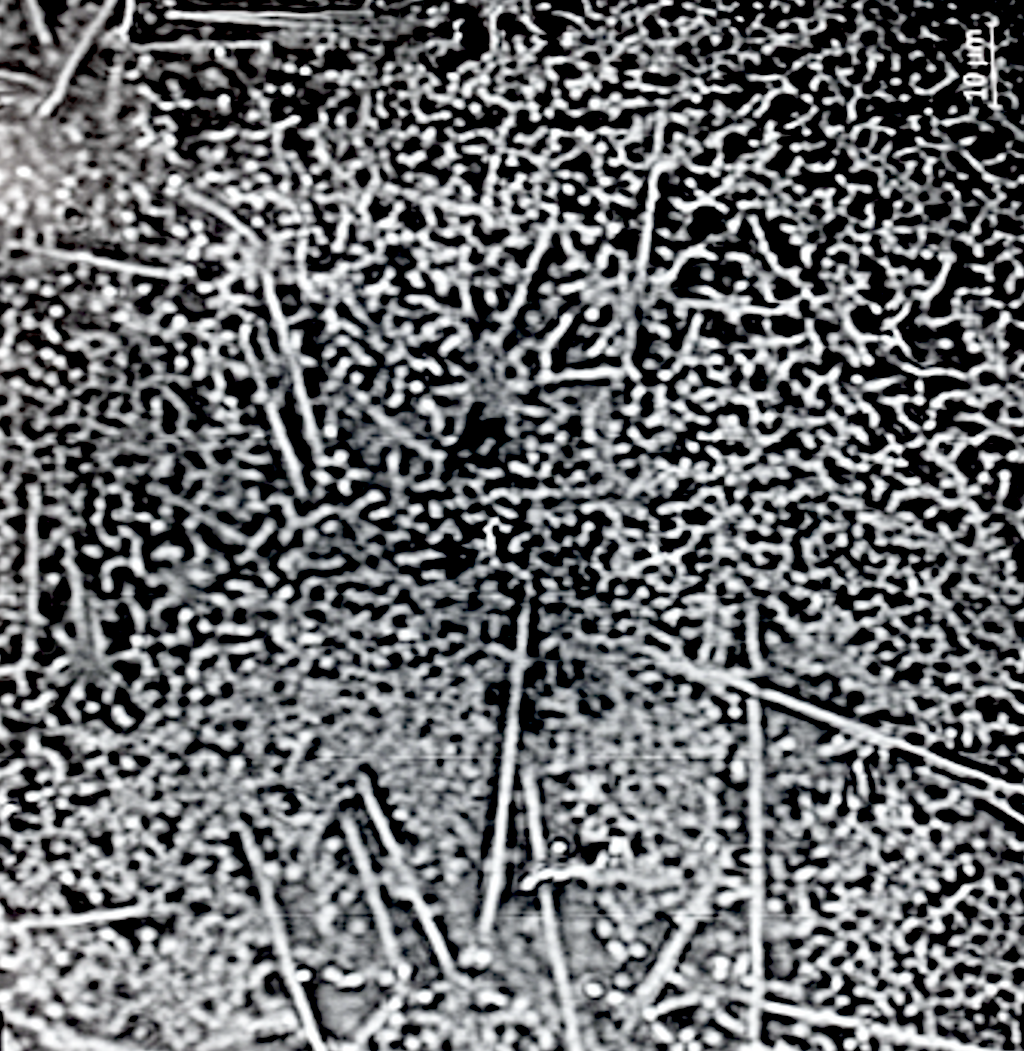}
    \caption{1000X optical image of reaction residue, showing dense, linear, and curdling formations.}
    \label{fig:optical_residue}
  \end{subfigure}
  \hspace{0.04\linewidth}
  \begin{subfigure}[t]{0.43\linewidth}
    \centering
    \includegraphics[height=2.5in, keepaspectratio]{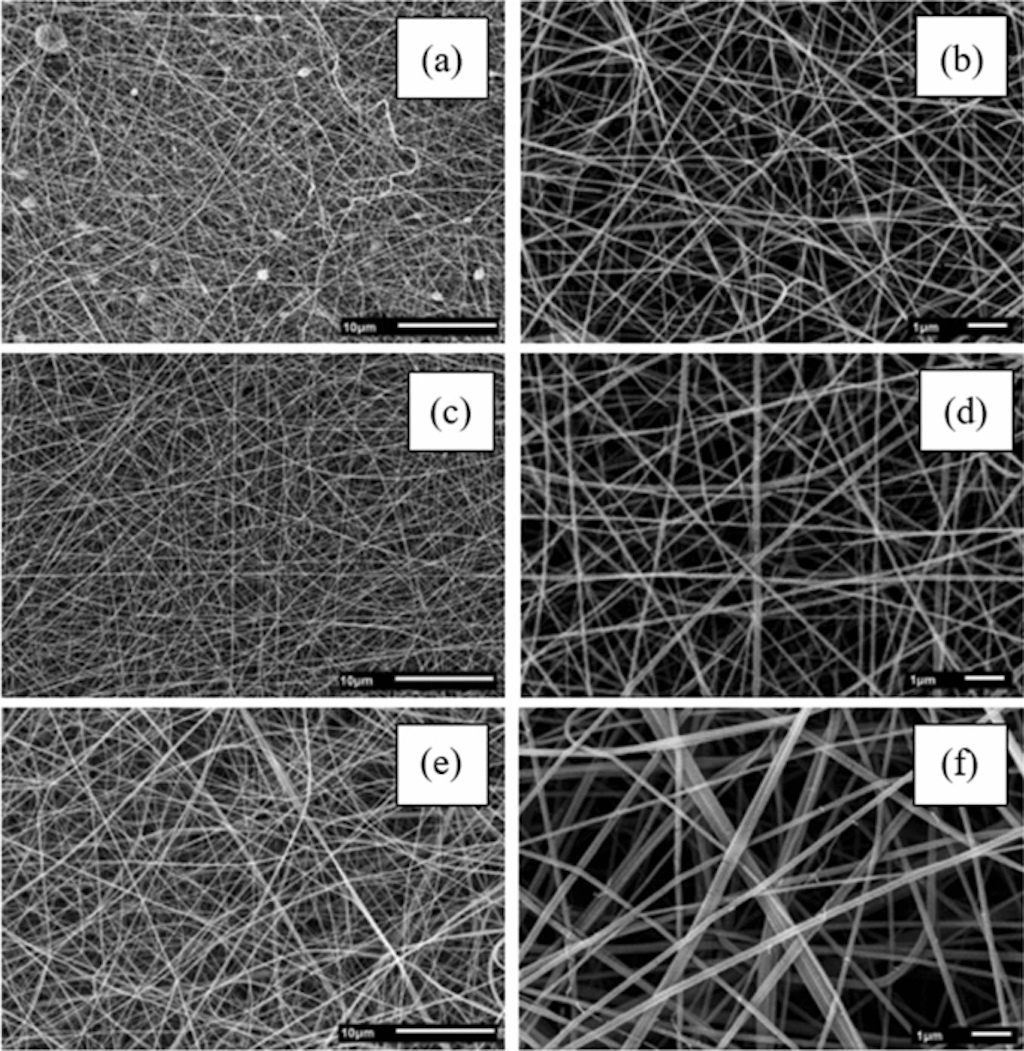}
    \caption{Comparative SEM image of gelatin-derived nanofibers from~\cite{ref07_farias2023}.}
    \label{fig:comparative_peptide}
  \end{subfigure}

  \caption{Morphological comparison of bubble-burst residue (left)\\ with peptide-like nanofiber structures observed in a Nature study (right).}
  \label{fig:fig4and5_combined}
\end{figure}

Microscopic analysis of the reactive residue provided deeper insight into the reaction products, revealing structural patterns indicative of complex molecular organization. At 1000X magnification, dense linear formations and twisted curdling patterns were observed (Figure~\ref{fig:fig4and5_combined}a, left), closely resembling the alignment and organization of peptide-like structures. A comparative analysis using SEM images from a 2023 Nature study~\cite{ref07_farias2023} provided a visual reference for evaluating morphological organization and nanofiber formation. This comparison (Figure~\ref{fig:fig4and5_combined}b, right) revealed striking similarities between the experimental formations and known gelatin-derived nanofibers, particularly in their folding and alignment patterns.

\begin{figure}[htbp]
  \centering

  \begin{subfigure}[t]{0.43\linewidth}
    \centering
    \includegraphics[height=3in, keepaspectratio]{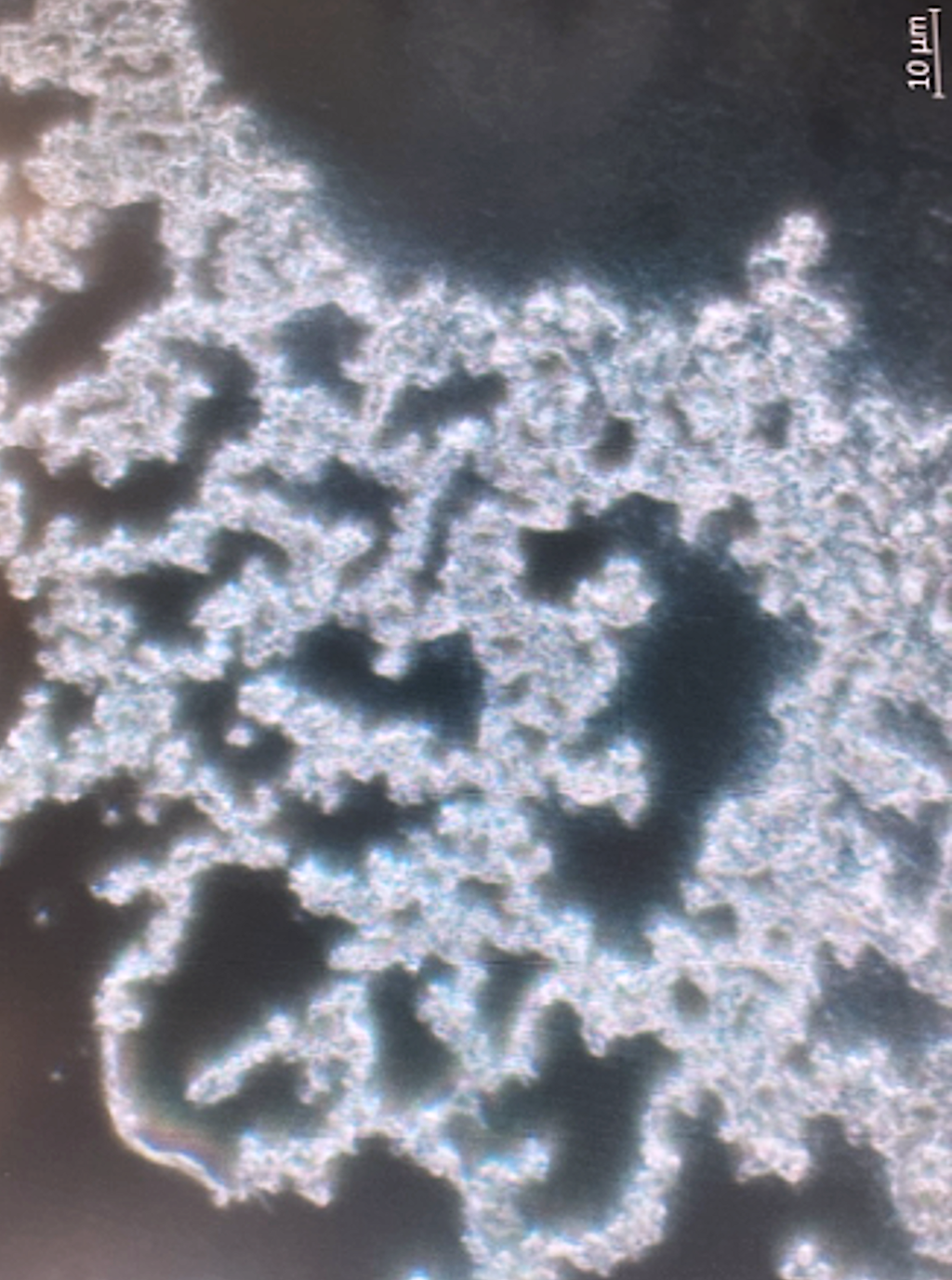}
    \caption{1000X dark-field image of reaction residue, revealing twisted and clustered peptide-like formations.}
    \label{fig:darkfield_peptide}
  \end{subfigure}
  \hspace{0.04\linewidth}
  \begin{subfigure}[t]{0.43\linewidth}
    \centering
    \includegraphics[height=3in, keepaspectratio]{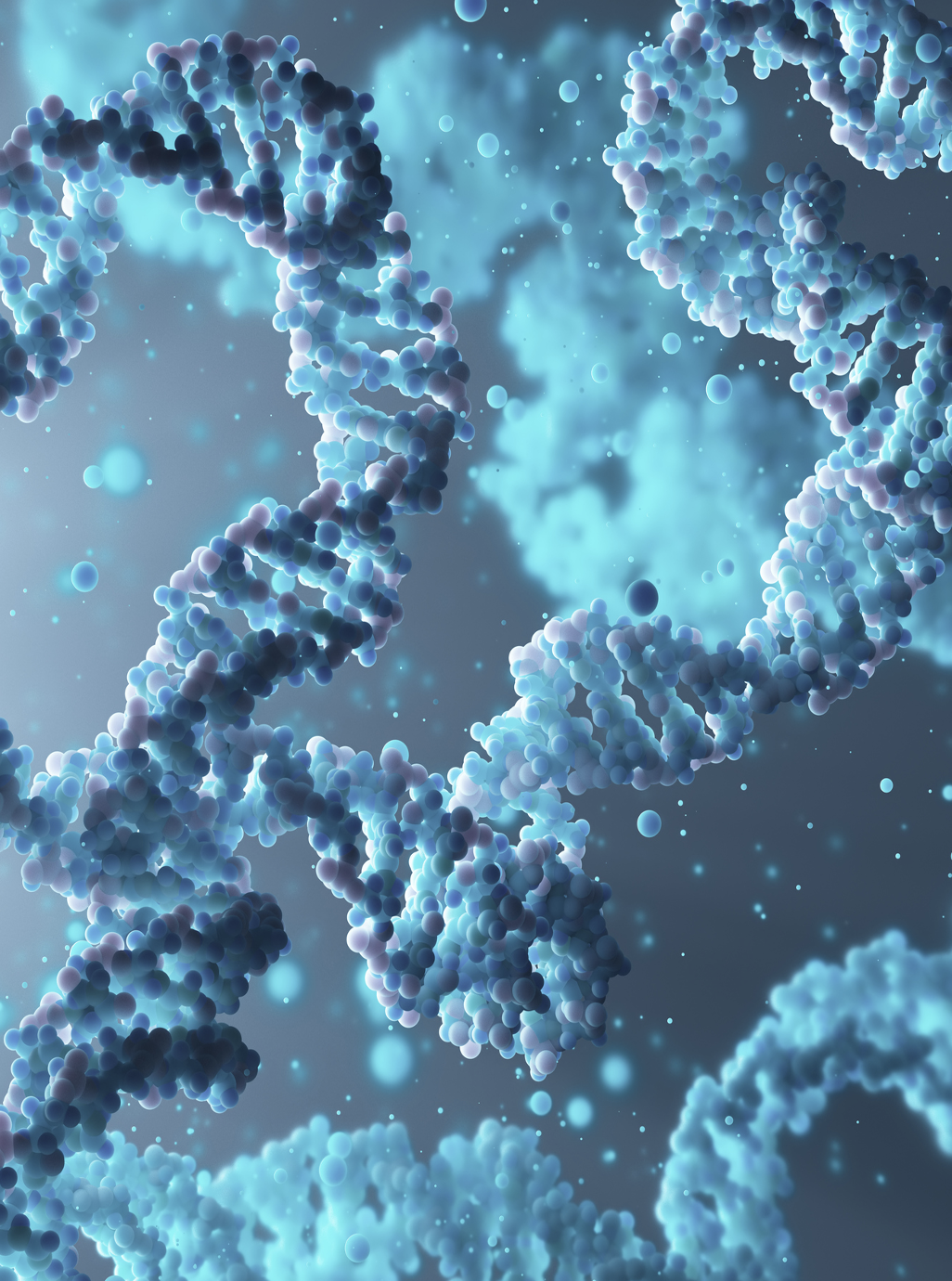}
    \caption{3D molecular illustration of a biomolecular chain (iStock~\cite{ref08_alllex2024}).}
    \label{fig:biomolecular_chain}
  \end{subfigure}

  \caption{Visual comparison of dark-field microscopy of bubble-burst residue (left)\\with a modeled biomolecular chain structure (right).}
  \label{fig:fig6and7_combined}
\end{figure}

Figure~\ref{fig:fig6and7_combined} continues the comparative analysis of microstructural features. A 1000X dark-field image of the reaction residue generated under identical bubble-bursting conditions in 30\% acetic acid reveals distinct particle-level organization characterized by twisted, folded, and entangled formations, suggestive of higher-order molecular assembly (Figure~\ref{fig:fig6and7_combined}a, left). A 3D molecular illustration provides a visual analog for interpreting these structures, highlighting the spatial complexity and self-organizing potential of peptide-like assemblies (Figure~\ref{fig:fig6and7_combined}b, right)~\cite{ref08_alllex2024}.

The consistent structural features observed across these microscopy and model-based comparisons provide compelling evidence that the bubble-bursting process fosters amino acid self-organization. These findings reinforce the broader hypothesis that localized high-energy microenvironments, generated under otherwise mild, catalyst-free conditions, are sufficient to initiate complex biochemical transformations that require ammonia or ammonium intermediates to form.

\section{Discussion}

With these results confirmed through multiple analytical techniques, the focus now turns to understanding the underlying mechanisms that drive these transformations, along with the broader implications of the bubble-bursting process for nitrogen-based chemistry.

\subsection{Bubble-Bursting Process}

\begin{figure}[H]
  \centering
  \includegraphics[width=0.48\linewidth]{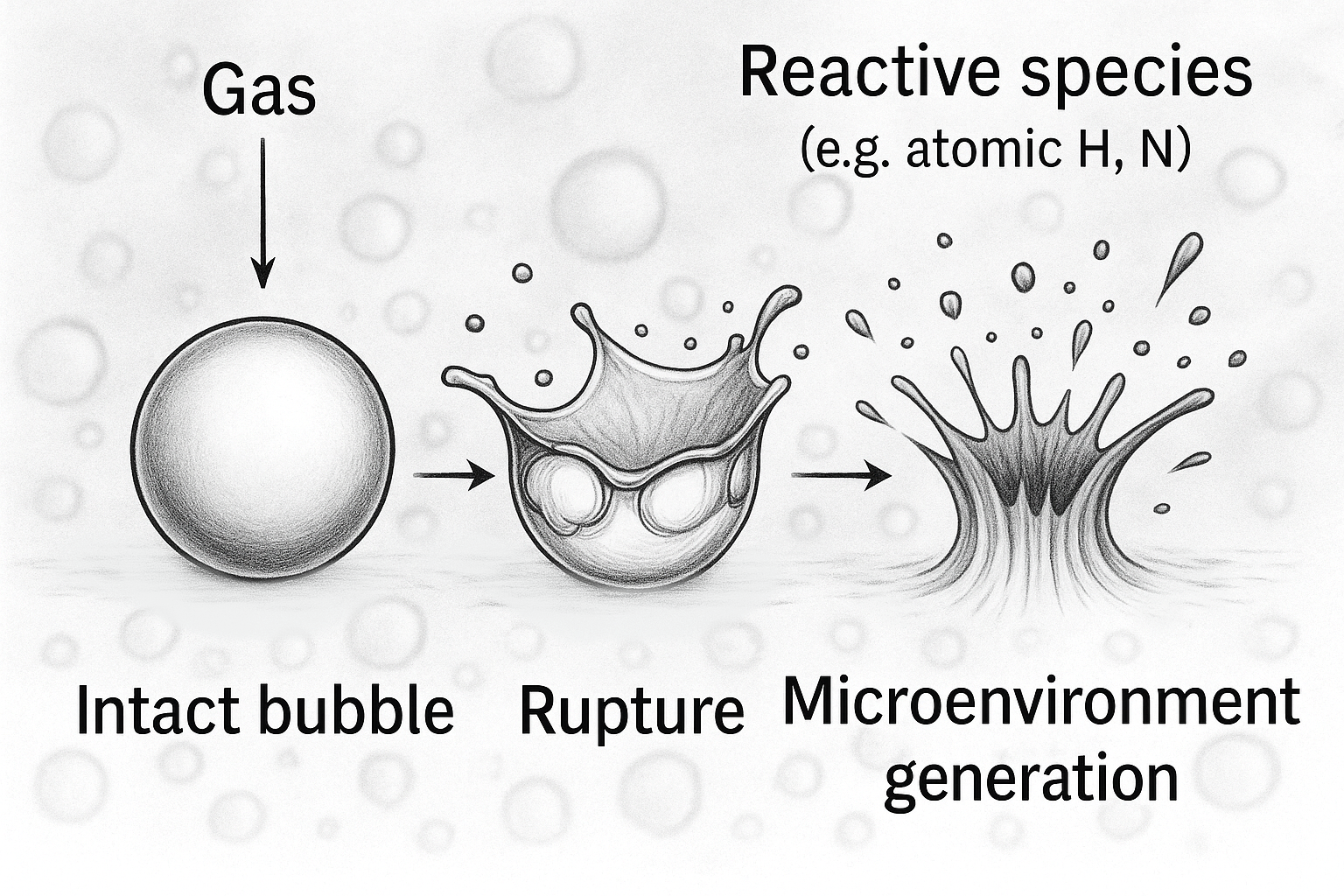}
  \caption{Bubble dynamics and microenvironment generation.}
  \label{fig:bubble_dynamics}
\end{figure}

The bubble-bursting process provides a simple yet effective means of driving chemical transformations under ambient conditions. When gas bubbles are introduced into an aqueous medium, they undergo cycles of expansion and collapse, culminating in rupture within the fluid. As conceptually illustrated in Figure~\ref{fig:bubble_dynamics}, these events release concentrated bursts of energy in the form of shockwaves, heat, and, in some cases, plasma.

Though brief and highly localized, these high-energy microenvironments are sufficient to overcome the activation barriers of stable molecules~\cite{ref09_lauterborn1997}. The dissociation of nitrogen (N\textsubscript{2}) and water (H\textsubscript{2}O), which possess strong covalent bonds, becomes energetically feasible under these transient conditions. The result is the formation of reactive intermediates that drive downstream chemical transformations.

Among the species generated, atomic hydrogen is particularly interesting. Molecular dynamics simulations and experimental studies of \href{https://www.sciencedirect.com/topics/materials-science/sonolysis}{sonolysis} have shown that water dissociation under similar energy conditions can produce atomic hydrogen, supporting its plausible role in enabling nitrogen activation and compound synthesis~\cite{ref10_penconi2015}.

In nitrogen fixation, atomic hydrogen lowers the activation barrier required for nitrogen (N\textsubscript{2}) dissociation, a crucial step in synthesizing nitrogen-based compounds. The highly energetic atomic hydrogen enables direct nitrogen–hydrogen interactions, unlike molecular hydrogen (H\textsubscript{2}), which is typically unreactive at standard temperature and pressure~\cite{ref11_threatt2022}. Given its electron-deficient nature, atomic hydrogen may act as a \href{https://chemed.chem.purdue.edu/genchem/topicreview/bp/ch11/lewis.php#:~:text=A%20Lewis%20acid%20is%20therefore,therefore%20an%20electron%2Dpair%20donor.}{Lewis acid}, readily accepting electrons from nitrogen or other reactants. This characteristic enhances its ability to form bonds and drive key chemical transformations~\cite{ref12_choi2020}.

Plasma formation at the bubble interface further amplifies the process, generating a diverse mix of high-energy reactive species. The high elongational strain rates produced during bubble bursts have been shown to shear microscopic particles and drive molecular transformations, supporting the role of bubble energy in facilitating these reactions~\cite{ref13_yu2015}.

The surrounding liquid medium stabilizes these reactive intermediates, allowing them to persist long enough to drive downstream chemical reactions essential for nitrogen-based compound formation. Atomic hydrogen is strongly suspected of playing a pivotal role in broader organic transformations beyond nitrogen fixation, including the synthesis of ammonia, amino acids, and urea, as explored in the following sections.

\subsection{Synthesis of Ammonia}

Ammonia synthesis, a cornerstone of nitrogen-based chemistry, is reimagined through the bubble-bursting process. This approach achieves instantaneous ammonia formation \textit{in situ} under ambient conditions, unlike the multi-step, energy-intensive Haber–Bosch method, which relies on high pressures, elevated temperatures, and complex catalysts~\cite{ref14_doe2024}.

In its most fundamental form, the system enables the direct formation of ammonium nitrate from atmospheric air (or nitrogen gas) and water. Under bubble-bursting conditions, reactive nitrogen species, such as nitrogen radicals (N\textsuperscript{*}), are generated through molecular dissociation. These species are thought to interact with atomic hydrogen (H\textsuperscript{*}), produced concurrently in the reaction medium, to form ammonia (NH\textsubscript{3}) \textit{in situ}~\cite{ref15_kozak2024}.

Following its formation, ammonia may react with inorganic anions, such as chloride, nitrate, sulfate, or phosphate, to form ammonium salts. The synthesized ammonia may also encounter oxidized nitrogen species such as nitrate (NO\textsubscript{3}\textsuperscript{--}) or nitrite (NO\textsubscript{2}\textsuperscript{--}), which can also arise from partial oxidation of nitrogen intermediates under aqueous, oxygen-rich conditions~\cite{ref16_zafiriou1987}. These products rapidly react within the aqueous medium to form ammonium nitrate (NH\textsubscript{4}NO\textsubscript{3}), a key nitrogen-containing compound with widespread agricultural relevance.

Supporting this hypothesis, recent research in conventional catalytic systems, such as those employing Ru/C12A7 electride, demonstrated that surface-adsorbed hydrogen plays a critical role in overcoming the kinetic barriers associated with nitrogen dissociation~\cite{ref17_kitano2018}. Similarly, the localized high-energy microenvironments produced during bubble bursts provide the activation energy necessary for these reactions to occur rapidly and efficiently. The overall reaction proceeds as follows:

\begin{equation}
\ce{N2 + 6H^{*} -> 2NH3}
\end{equation}

In addition, ammonia may engage in reactions with other compounds in solution, producing valuable ammonium derivatives. For instance, in the presence of dissolved carbon dioxide (CO\textsubscript{2}), ammonia forms ammonium carbonate through the following reaction:

\begin{equation}
\ce{NH3 + CO2 -> NH2COONH4}
\end{equation}

Similarly, when hydrochloric acid (HCl) is introduced, ammonia reacts to form ammonium chloride:

\begin{equation}
\ce{NH3 + HCl -> NH4Cl}
\end{equation}

This observation aligns with findings from a recently proposed method involving freely oscillating microbubbles, where controlled dynamics under adiabatic conditions generate sufficient internal energy to enable nitrogen activation. Although this model relies on sustained oscillation rather than rupture and collapse, the shared localized energy profiles offer theoretical support for nitrogen dissociation observed in the bubble-bursting process~\cite{ref18_kovacs2024}.

\subsection{Synthesis of Amino Acids}

With ammonia readily accessible as a reactive intermediate, the process for synthesizing more intricate nitrogen-based compounds becomes feasible. The spontaneous formation of amino acids—the essential building blocks of life—becomes possible under the same high-energy microenvironments that drive nitrogen fixation. Within these intense, transient reaction zones, ammonia interacts with organic precursors to synthesize amino acids. This transformation is significant because it demonstrates that the bubble-bursting process not only generates reactive nitrogen species but also enables sequential chemical steps necessary for biomolecular synthesis. Ammonia's interaction with carbon-containing compounds under energy-rich conditions facilitates amino acid formation, paralleling known abiotic synthesis pathways without requiring enzymatic catalysis or external catalysts.

For instance, in a solution containing acetic acid (\ce{CH3COOH}), ammonia (\ce{NH3}) reacts to produce glycine (\ce{NH2CH2COOH}), the simplest amino acid, illustrating a potential pathway for amino acid formation. This transformation is believed to occur as the amine group (\ce{-NH2}) from ammonia bonds to the carbon adjacent to the carboxyl group (\ce{-COOH}) in acetic acid, following the reaction:

\begin{equation}
\ce{CH3COOH + NH3 -> NH2CH2COOH}
\end{equation}

The system's versatility extends beyond glycine, enabling the synthesis of a broader range of amino acids. It can also facilitate peptide formation by introducing additional reactants or modifying the reaction medium, further expanding its potential for complex biochemical synthesis. For example, amino acids such as alanine or cysteine may be accessed by introducing appropriate precursors. In the case of alanine, this can be achieved via the \href{https://www.masterorganicchemistry.com/2018/11/12/the-strecker-synthesis-of-amino-acids/}{Strecker-type pathway} using acetaldehyde as a carbonyl precursor—a route well established in prebiotic chemistry models~\cite{ref19_powner2009}.

The ability of bubble-burst phenomena to drive amino acid and peptide formation is further supported by studies demonstrating peptide synthesis from amino acids using bubble bursting combined with arc plasma~\cite{ref20_wei2024}. Unlike arc plasma methods that require external high-energy inputs, the bubble-bursting process described here generates localized high-energy microenvironments naturally, offering a more straightforward and energy-efficient approach.

\subsection{Synthesis of Urea}

This system presents new opportunities for synthesizing additional nitrogen-based compounds, most notably urea (\ce{CO(NH2)2}). In conventional chemistry, urea formation involves a well-established two-step reaction: ammonia (\ce{NH3}) reacts with dissolved carbon dioxide (\ce{CO2}) to produce ammonium carbamate (\ce{NH2COONH4}):

\begin{equation}
\ce{NH3 + CO2 -> NH2COONH4}
\end{equation}

The subsequent dehydration of ammonium carbamate yields urea:

\begin{equation}
\ce{NH2COONH4 -> CO(NH2)2}
\end{equation}

While the first step is exothermic, the second typically requires elevated temperatures and pressures to overcome kinetic barriers~\cite{ref21_ding2023}. In contrast, and as confirmed by Raman spectroscopy, this bubble-bursting method generates urea directly and under ambient conditions, without added energy or prolonged reaction time. Both ammonia and its downstream products form, react, and remain entirely within the reaction medium, rather than being introduced or extracted as separate inputs.

The presence of urea suggests that both reaction steps may occur near-instantaneously (on the order of femtoseconds) as part of a unified energy event driven by bubble dynamics. While the mechanism is not yet fully understood, the data imply that the reactive microenvironments generated by this process may be sufficient to drive not only ammonia formation but its rapid transformation into higher-order compounds. This opens new possibilities for decentralized, on-demand urea production, circumventing the limitations of conventional thermal pathways.

Different elements, such as sulfur and phosphorus, can be incorporated into solution via sulfuric acid, phosphoric acid, nitric acid, or hydrochloric acid. These additions further expand the versatility of the bubble-bursting process, facilitating the formation of fertilizers enriched with modified amino compounds and other nitrogen-based structures. While some of these outcomes have been validated in laboratory experiments, the corresponding data are not included in this report. This approach broadens the potential applications of bubble-burst synthesis across biochemical, agricultural, and industrial domains.

\subsection{System Parameters and Enhancements}

Several system parameters may be adjusted to influence product outcomes in addition to the core reaction conditions. The gas feed may include atmospheric air, purified molecular nitrogen (\ce{N2}), or mixtures containing trace nitrogen compounds. The system is also compatible with isotopic variants, such as nitrogen-15 (\ce{\textsuperscript{15}N2}), carbon-13 (\ce{\textsuperscript{13}CO2}), and carbon-14 (\ce{\textsuperscript{14}CO2}), enabling labeled product synthesis for applications in diagnostics, research, and metabolic tracing. These variations do not alter the mechanics of the core reaction, but may influence the handling, monitoring strategies, or purification requirements of the downstream product.

Beyond its efficiency, this method offers a sustainable pathway for synthesis of nitrogen-based compounds by integrating with carbon capture systems, converting \ce{CO2} into agriculturally and industrially essential products. This dual benefit, sequestering carbon while producing critical compounds such as urea, presents a scalable approach to reducing emissions and improving global food security. The process transforms waste carbon into a valuable resource while simplifying production, eliminating multi-stage conversions and large-scale facility requirements, and enabling decentralized, on-demand manufacturing.

\section{Conclusion}

Bubble bursting is a well-documented natural phenomenon, particularly in ocean environments, where it contributes to sea spray formation and the release of organic and inorganic compounds into the atmosphere~\cite{ref22_wikipedia_seaspray}. Several scientific methods referenced in this paper have sought to replicate aspects of this process, particularly in aerosol physics, mechanochemistry, and microbubble technology. The present method stands out as one of the most straightforward and direct means of harnessing bubble-bursting to drive nitrogen transformations in a controlled manner.

Given that the synthesized products are water-soluble and naturally integrated within the reaction medium, they are immediately suited for direct application—an advantage that simplifies downstream processing. In agriculture, the system's ability to synthesize fertilizers such as ammonium compounds and urea directly on-site offers significant benefits. Large-scale hydroponic farms can benefit from a continuous, sustainable supply of tailored fertilizers, while traditional farms can apply the solution directly through fertigation. Compact versions of the system may also serve home gardeners with environmentally friendly, customizable fertilizer production.

The environmental benefits of the bubble-bursting process extend beyond energy efficiency. Integration with carbon capture technologies offers a pathway for converting captured greenhouse gases into valuable nitrogen-based products. This dual-purpose approach aligns with global sustainability goals by combining \ce{CO2} utilization with chemical synthesis, thereby reducing emissions while generating high-value outputs.

Experimental results demonstrate that bubble-bursting can convert basic inputs—nitrogen gas, carbon dioxide, and water—into valuable nitrogen-based compounds, including ammonia, amino acids, and urea. This transformation is enabled by localized high-energy microenvironments created during bubble collapse and rupture, which initiate molecular dissociation and drive the formation of reactive intermediates such as atomic hydrogen. Operating under ambient conditions and without the need for catalysts or external energy sources, this method contrasts sharply with energy-intensive approaches like the Haber–Bosch process or arc plasma systems.

Its modularity and simplicity make the system well-suited for decentralized chemical production. With minimal infrastructure and scalable inputs, the bubble-bursting method offers a practical and adaptable platform for applications across agriculture, pharmaceuticals, and industrial chemistry.

\section*{Author Contributions}

\href{https://mse.ncsu.edu/people/cuomo/}{Jerome J. Cuomo} conceived the bubble-burst process and served as principal investigator, leading the theoretical development, experimental work, and scientific direction. A \href{https://clintonwhitehouse4.archives.gov/Initiatives/Millennium/capsule/cuomo.html}{1995 National Medal of Technology} recipient, Dr. Cuomo has made pioneering contributions to materials science through innovations in plasma processes, synthesis, and engineering applications.

\href{https://www.linkedin.com/in/ian-goodall-aaaa17189/}{Ian Goodall}, a graduate student in the Department of Materials Science and Engineering at North Carolina State University, performed Raman spectroscopy and chemical analysis. His efforts were instrumental in identifying and confirming urea and other nitrogen-based compounds, as well as supporting the review and editing of this manuscript.

\href{https://mse.ncsu.edu/people/rguarni/}{C. Richard Guarnieri}, Adjunct Associate Professor at North Carolina State University, contributed subject matter expertise and participated in technical reviews. His background in physical chemistry and catalysis guided experimental design and validation of theoretical aspects.

\href{https://www.linkedin.com/in/stephenhudak/}{Stephen Hudak}, Materials Science Engineer, contributed to Raman analysis and conceptual prototyping, and provided insight into possible chemical transformations occurring in the bubble-bursting system.

\href{https://en.wikipedia.org/wiki/Jerry_Cuomo}{Gennaro (Jerry) Cuomo}, Retired IBM Fellow, Founder of \href{https://wildducks.us}{Wild Ducks LLC}, and Lecturer at NC State University, served as the principal writer. He collaborated with his dad, Jerome J. Cuomo, on manuscript structure, figures, references, conceptual synthesis, and refinement for clarity and flow.

\section*{Intellectual Property Notice}
The methods and systems described in this manuscript are the subject of U.S. Patent Application No. 19/183,880. Academic citation and discussion are welcomed. For all other inquiries please contact Jerome J. Cuomo.

%Bibliography
%\bibliographystyle{unsrt}  
%\bibliography{Bubble_Burst_Synthesis_Ammonia}  

\end{document}